\documentclass[osajnl,twocolumn,showpacs,superscriptaddress,10pt]{revtex4-1} %% use 11pt for Applied Optics
\usepackage{amsmath,amssymb,graphicx}
\begin{document}

\title{Spiraling elliptic breathers in saturable nonlinear media with linear anisotropy}

\author{Guo Liang}
\affiliation{Laboratory of Nanophotonic Functional Materials and Devices, South
China Normal University, Guangzhou 510631}

\author{Qi Guo}\email{Corresponding author: guoq@scnu.edu.cn}
\affiliation{Laboratory of Nanophotonic Functional Materials and Devices, South
China Normal University, Guangzhou 510631}

\author{Qian Shou}
\affiliation{Laboratory of Nanophotonic Functional Materials and Devices, South
China Normal University, Guangzhou 510631}

\author{Zhanmei Ren}
\affiliation{Laboratory of Nanophotonic Functional Materials and Devices, South
China Normal University, Guangzhou 510631}

\begin{abstract}
We have introduced a class of spiraling elliptic breathers in saturable nonlinear media with linear anisotropy. Two kinds of evolution behaviors of the breathers, rotating and pendulum-like librating, are both predicted by the variational approach, and confirmed by the numerical simulation.The spiraling elliptic breathers can rotate even though they have no initial orbital angular momentum (OAM). Due to the linear anisotropy of the media, the OAM is no longer conserved. Therefore, the angular velocity is found to be not a constant but a periodic function of the propagation distance. When the linear anisotropy is large enough, the spiraling elliptic breathers can librate like the pendulum. The spiraling elliptic breathers exist in the media with not only the saturable nonlinearity but also the nonlocal nonlinearity, as a matter of fact, they are universal in the nonlinear media with the linear anisotropy.
\end{abstract}

% REPLACE WITH CORRECT OCIS CODES FOR YOUR ARTICLE
                          % NOTE: \ocis{} IS ALIASED TO \pacs{} BUT MUST
                          % FORMAT THE TERMS CORRECTLY FOR EACH JOURNAL

\maketitle %% required

\section{Introduction}

The self-trapping and self-focusing of optical beams in nonlinear media of both linear isotropy (isotropic diffraction) and
nonlinear isotropy have been studied extensively for over three decades~\cite{trillo-book-2001,kivshar-book-2003,shen-book-1984}. In an isotropic medium, the circular symmetry is conserved, and therefore the fundamental solitons have circularly-symmetrical shape. Introducing the nonlinear anisotropy into the medium, the self-trapping beams with
ellipse-shaped spots can be obtained. Coherent elliptic strongly nonlocal solitons were observed
experimentally in lead glass~\cite{Rotschild-prl-05}, where the nonlinear
anisotropy is achieved by rectangular boundaries in the transverse. Elliptical discrete solitons can
form in an optically induced two-dimensional photonic lattice, where
the nonlinear anisotropy comes of enhanced photorefractive
anisotropy and nonlocality under a nonconventional bias
condition~\cite{Zhang-oe-08}. It was reported recently that the anisotropic nonlocal nonlinearity of the diffusive (thermal) type can stabilize the dipole-mode solitons, which are completely unstable in the isotropic medium~\cite{ye-pra-2010}.
%It was
%predicted very recently~\cite{Desyatnikov-prl-10} that the
%elliptic solitons with the initial orbital angular momentum (OAM)
%can exist in isotropic medium, and they will rotate along the propagate
%distance.

%There are few literatures to report the self-trapping and self-focusing of a optical beam in nonlinear media of linear anisotropy, except for Refs.\cite{guo-joa-1999,polyakov-pre-2002,conti-pre-2005} to our knowledge.

The linear anisotropy is also important in many soliton phenomena. In Ref.\cite{guo-joa-1999}, the self-focusing of the
beam propagating in any direction in uniaxial crystals was discussed, there exists an elliptic
self-trapping beam for the extraordinary light in uniaxial crystals. Stationary elliptic quadratic solitons~\cite{polyakov-pre-2002} and elliptic nonlocal solitons~\cite{conti-pre-2005} were found successively in biaxial crystals and the nematic liquid crystals with large birefringence, respectively. The generation of multiple solitons~\cite{polyakov-ol-2002,Malendevich-ol-2002} were reported theoretically and experimentally to be the consequence of the linear anisotropy. The linear isotropy can also make the elliptic optical beams without the initial orbital angular momentum (OAM) rotate during the linear propagation~\cite{chen-oc-2011}, and the rotation angle will monotonously approach to the value determined by the media and the input parameters of the beam. It was
predicted very recently that the spiraling
elliptic solitons with the OAM
can exist in isotropic saturable nonlinear media~\cite{Desyatnikov-prl-10} and in isotropic nonlocal nonlinear media~\cite{liang-pra-2013}, where the OAM can bring in the effective linear anisotropy.
The nonlinear propagation of elliptic optical beams in saturable nonlinear medium with the linear anisotropy is investigated in the paper.
A class of spiraling elliptic breathers is found to exist in such media, which can rotate even though
they have no initial OAM. The angular velocity is no longer a constant but a periodic function of the propagation distance. The libration of the spiraling elliptic breathers like the pendulum is discovered when the linear anisotropy of the media is large enough.

\section{The variational solution of the spiraling elliptic breathers}
\begin{figure}[htb]
\centerline{\includegraphics[width=3.0cm]{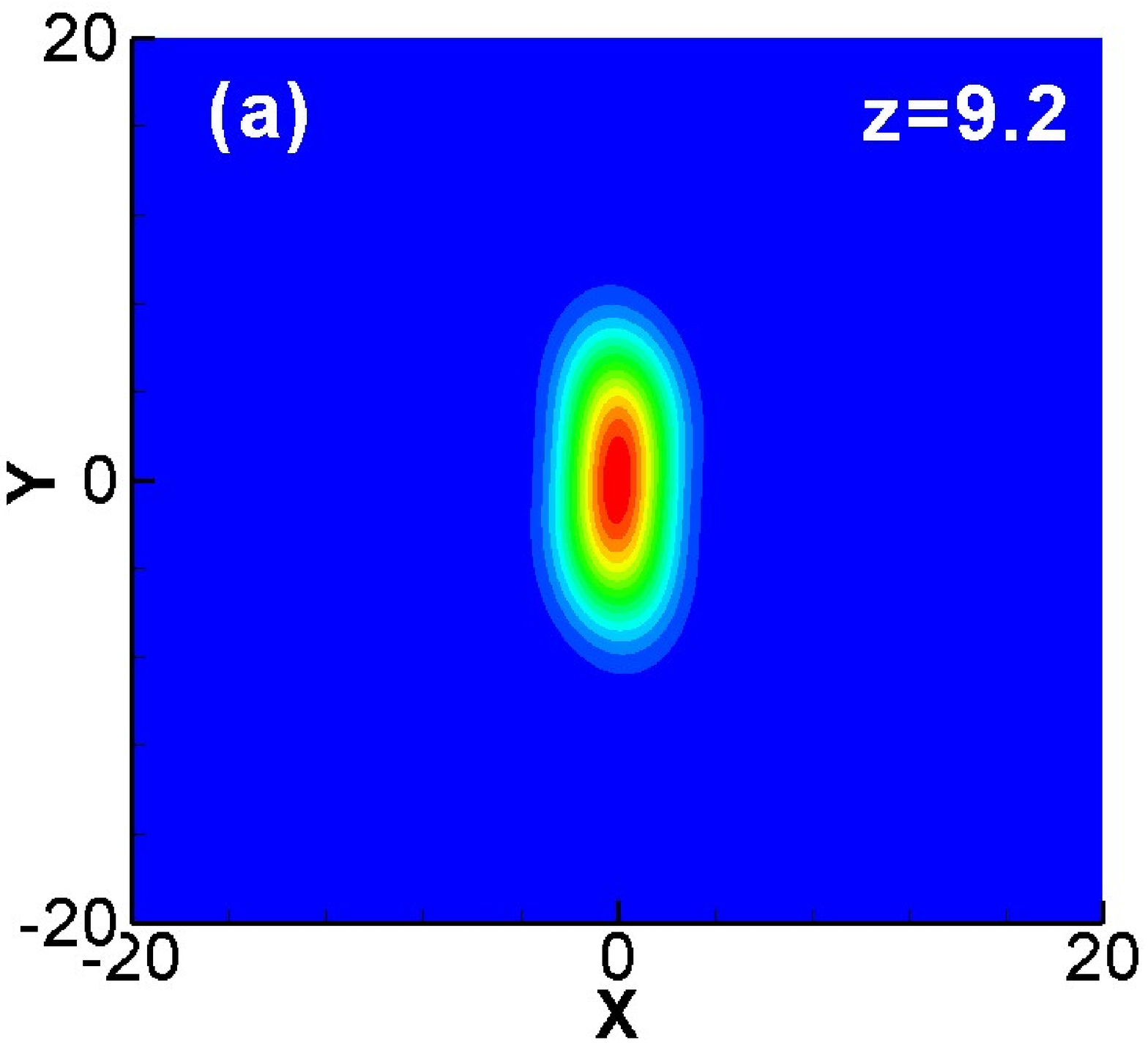}
\includegraphics[width=3.0cm]{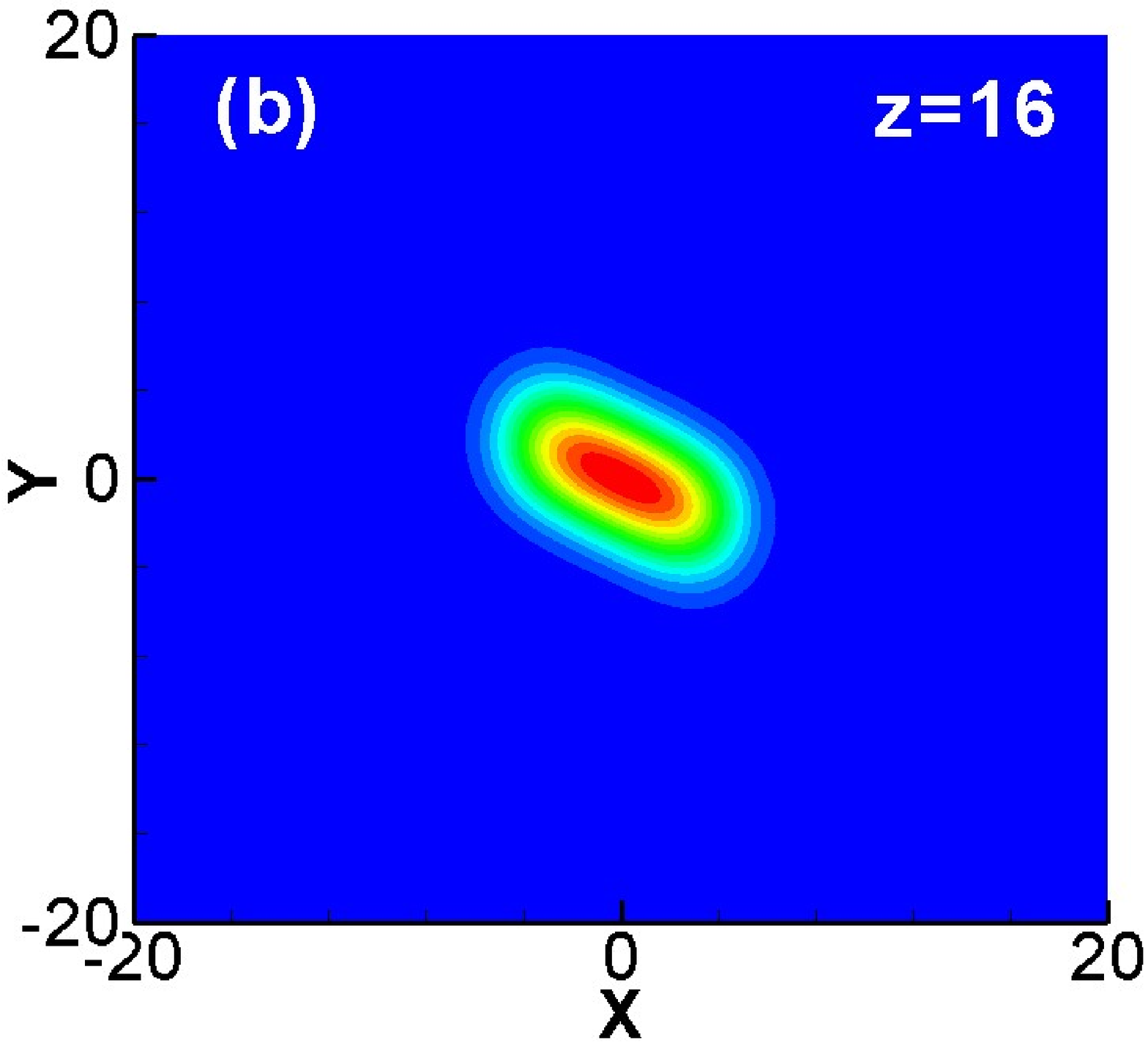}}
\centerline{
\includegraphics[width=3.0cm]{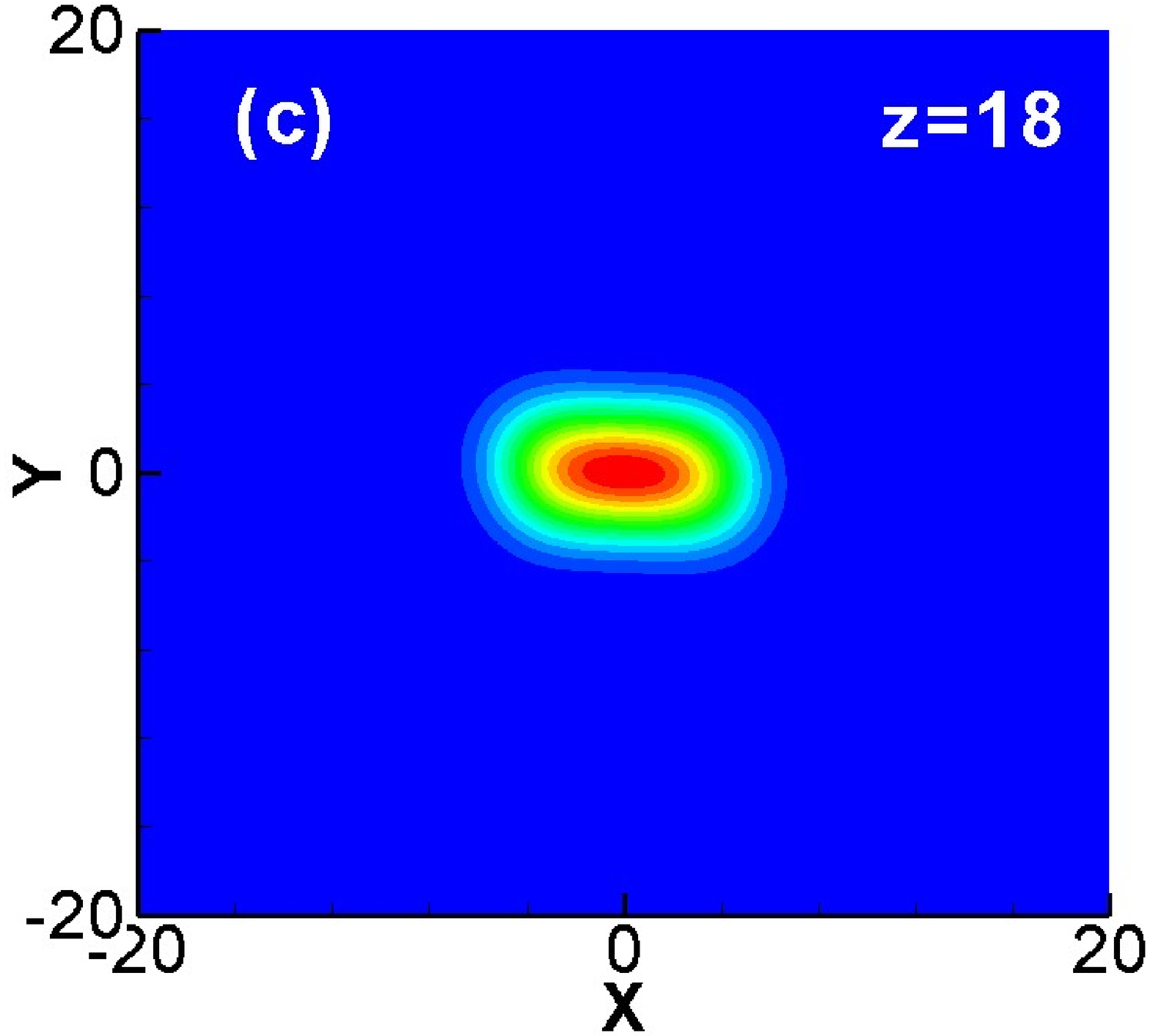}
\includegraphics[width=3.0cm]{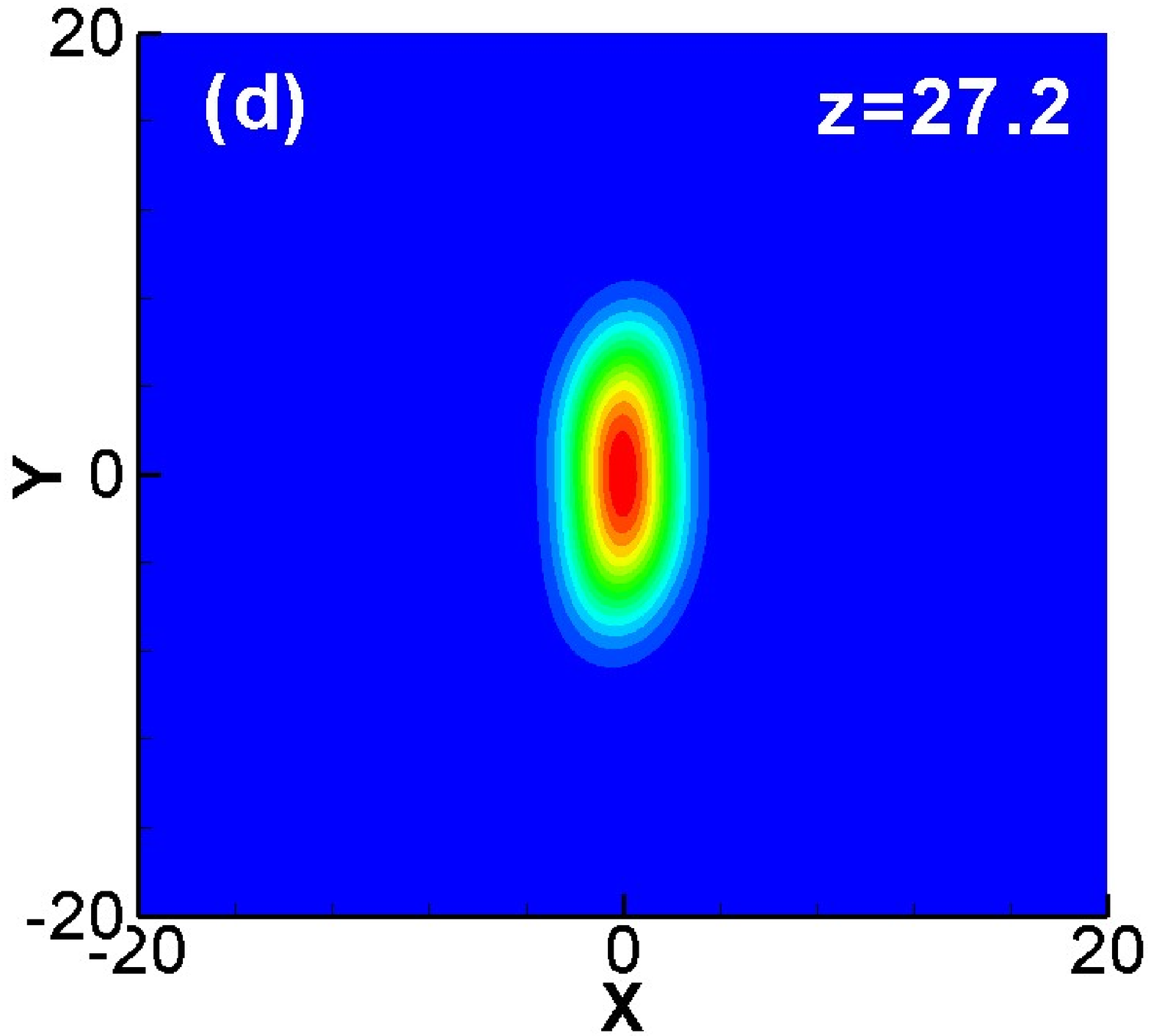}}
\caption{(color online) Propagation of the spiralling optical breather. The parameters are $\alpha_1=1.0,\alpha_2=1.3,b=4.26,c=2.13,P=127.32,\Theta=0.052,\beta_0=0 $ and $\theta=0.$}\label{across}
\end{figure}

\begin{figure}[htb]
\centerline{\includegraphics[width=3.5cm]{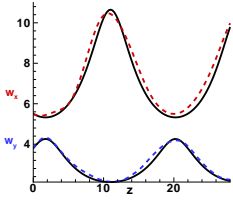}}
\caption{(color online) Comparison of the beam width of the
analytical solution (black solid curves) with that of the numerical simulation (red dashed
curves for $w_x$ and blue dashed curves for $w_y$). The parameters are the same as those in figure~\ref{across}.}\label{width}
\end{figure}

The propagation of optical beams in saturable nonlinear media with linear anisotropy can be modeled by the following nonlinear Schr\"{o}dinger equation (NLSE)~\cite{kivshar-book-2003,guo-joa-1999}
\begin{equation}\label{NLSE-dimensional}
i\frac{\partial A}{\partial\zeta}+\frac{1}{2k}\left(\alpha_1^2\frac{\partial^2A}{\partial\xi^2}+\alpha_2^2\frac{\partial^2A}{\partial\eta^2}\right)+\frac{kn_2}{n_0}\frac{I}{1+I/I_s}A=0,
\end{equation}
where $A(\xi,\eta,\zeta)$ is a paraxial beam, $I=|A|^2$ is the beam intensity, $I_s$ is the saturation intensity, $n_2$ is the nonlinear index coefficient, $\zeta$ is the longitudinal
coordinate, $\xi$ and $\eta$ are the transverse coordinates, $k$ is the wavenumber in the media without nonlinearity, $n_0$ is the linear refractive index of the media, $\alpha_1$ and $a_2$ are the diffraction coefficients along $\xi$ and $\eta$ directions, respectively. The lager the coefficients are, the more strongly the optical beam diffracts in those directions. When $\alpha_1=\alpha_2$, an optical beam of the circular shape will
diffract equally along any direction~\cite{liang-pra-2013}. But here we consider the case of linear anisotropy, i.e., $\alpha_1\neq\alpha_2$. Introducing the scaled dimensionless variables as $x=\xi/w_0,y=\eta/w_0,z=\zeta/L_d, \varphi=\left(\frac{kn_2L_d}{n_0}\right)^{1/2}A$, where $L_d=2kw_0^2$ is the Rayleigh distance, equation~(\ref{NLSE-dimensional}) takes the form of
\begin{equation}\label{NSE1}
  i\frac{\partial\varphi}{\partial z}+\alpha_1^2\frac{\partial^2\varphi}{\partial x^2}+\alpha_2^2\frac{\partial^2\varphi}{\partial y^2}+\frac{|\varphi|^2}{1+|\varphi|^2/\gamma}\varphi=0,
\end{equation}
where $\gamma$ is the dimensionless saturation intensity, and $\gamma=1$ is assumed in the paper.
 Through the variable transformation $X=x/\alpha_1,Y=y/\alpha_2,Z=z$, the NLSE (\ref{NSE1}) becomes
\begin{equation}\label{NSE2}
  i\frac{\partial\varphi}{\partial Z}+\frac{\partial^2\varphi}{\partial X^2}+\frac{\partial^2\varphi}{\partial Y^2}+\frac{|\varphi|^2}{1+|\varphi|^2}\varphi=0,
\end{equation}
where the optical beam is changed to $\varphi(X,Y,Z)$. The Lagrangian of equation~(\ref{NSE2}) can be expressed
as~\cite{Desyatnikov-prl-10} $L=i/2\int\!\!\!\int(\varphi^*\partial
\varphi/\partial Z-\varphi\partial \varphi^*/\partial Z){\rm d}X{\rm d}Y-H, $
where $H$ is the Hamiltonian of the system,
\begin{equation}\label{Hamiltonianex}
 H=\int\!\!\!\int\left(\left|\frac{\partial\varphi}{\partial X}\right|^2+\left|\frac{\partial\varphi}{\partial Y}\right|^2-|\varphi|^2+\ln(1+|\varphi|^2)\right){\rm d}X{\rm d}Y.
\end{equation}
We introduce a trial
function~\cite{Desyatnikov-prl-10},
\begin{equation}\label{trial}
\varphi=\sqrt{\frac{P}{\pi
b(Z)c(Z)}}G\left[\frac{\xi}{b(Z)}\right]G\left[\frac{\eta}{c(Z)}\right]\exp(i\phi),
\end{equation}
where the Gaussian envelope is $G(t)=\exp(-t^2/2)$, the phase is
$\phi=B(Z)\xi^2+\Theta(Z)\xi\eta+Q(Z)\eta^2+\theta(Z)$, $\xi=X\cos
\beta(Z)+Y\sin\beta(Z),\eta=-X\sin\beta(Z)+Y\cos\beta(Z)$ and $P=\int\!\!\!\int|\varphi|^2{\rm d}X{\rm d}Y$ is
the power. From equation~(\ref{trial}), we can obtain
the OAM,
$M=\mbox{Im}\int\!\!\!\int\varphi^*(X\frac{\partial\varphi}{\partial Y}-Y\frac{\partial\varphi}{\partial X}){\rm d}X{\rm d}Y=1/2P(b^2-c^2)\Theta$. Inserting the trial solution
(\ref{trial}) into the Lagrangian, $L$ can be analytically
determined. By using the variational approach~\cite{Anderson-pra-83} we can obtain that
$P'=0, H'=0, M'=0, \beta'=2(b^2+c^2)\Theta/(b^2-c^2)$ and
$
 b'=4bB, c'=4cQ,
$
where the primes indicate derivatives with respect to the variable
$Z$. Inserting relational expressions above into the Hamiltonian (\ref{Hamiltonianex}), we obtain
\begin{equation}\label{Hamiltonian}
  H=\frac{P}{8}\left(b'^2+c'^2\right)+\Pi,
\end{equation}
where $\Pi=\frac{P}{2}\left[\frac{1}{b^2}+\frac{1}{c^2}+4\sigma^2\frac{b^2+c^2}{(b^2-c^2)^2}\right]-\pi bcLi_2\left(-\frac{P}{\pi bc}\right)-P$ is the potential of the system with $\sigma=M/P$. Here, $Li_2(t)$ is the dilogarithm function, defined by $Li_2(t)=\int_t^0dq\ln(1-q)/q$.
Solitons are corresponding to the extremum of the potential $\Pi(b,c)$. So letting $\partial\Pi/\partial b=\partial\Pi/\partial c=0$, we obtain
\begin{eqnarray}
  \frac{4}{b^3}+\frac{16 b\sigma ^2 \left(b^2+3c^2\right) }{\left(b^2-c^2\right)^3}+\frac{4\pi c}{P}F(b,c)&=& 0, \label{1}\\
  \frac{4}{c^3}-\frac{16 c \sigma ^2\left(3b^2+c^2\right) }{\left(b^2-c^2\right)^3}+\frac{4\pi b}{P}F(b,c) &=& 0,\label{2}
\end{eqnarray}
where $F(b,c)=\left[\ln\left(1+\frac{P}{\pi bc}\right)+Li_2\left(-\frac{P}{\pi bc}\right)\right]$. If $b$ and $c$ are given, the critical power and the critical OAM can be obtained from equations~(\ref{1}) and (\ref{2}), with which the optical beam can propagate keeping its elliptic profile changeless and rotating stably in the $XYZ$-coordinate frame. The rotational angular velocity can be obtained as
\begin{equation}\label{angular velocity}
\omega=\beta'=\frac{4\sigma(b^2+c^2)}{(b^2-c^2)^2}.
\end{equation}
Then, we can obtain $\beta(Z)=\omega Z+\beta_0$, where $\beta_0$ is the initial inclination of the elliptic optical beam at $Z=0$ in the $XYZ$-coordinate frame. Here, we give an example that~\cite{Desyatnikov-prl-10}, when $b=4.26,c=2.13$ the critical power $P=127.32\pi$, the critical OAM $\sigma=0.35$, and the rotational angular velocity $\omega=0.17$. The optical beam expressed in the $xyz$-coordinate frame (the laboratory frame) is of the form
{\setlength\arraycolsep{0pt}
\begin{eqnarray}
  \varphi&=&\sqrt{\frac{P}{\pi bc}}\exp\left[-\frac{(\frac{x\cos\beta}{\alpha_1}+\frac{y\sin\beta}{\alpha_2})^2}{2b^2}-\frac{(\frac{y\cos\beta}{\alpha_2}-\frac{x\sin\beta}{\alpha_1})^2}{2c^2}\right]\nonumber\\
  &\times&\exp\left[i\Theta(\frac{x\cos\beta}{\alpha_1}+\frac{y\sin\beta}{\alpha_2})(\frac{y\cos\beta}{\alpha_2}-\frac{x\sin\beta}{\alpha_1})+i\theta\right].\label{solution}
\end{eqnarray}}
The evolution of the optical beam in the $xyz$ coordinate system, as an example, is shown in figure~\ref{across}, where the input beam is expressed as equation~(\ref{solution}) and $\alpha_1=1.0,\alpha_2=1.3$. The optical beam rotates during the propagation, while its shape changes periodically. To confirm the validity of the approximately analytical solution, we compare the two half widths obtained from
variational solution, $w_x=\alpha_1(b^{-2} \cos^2\omega_c z+c^{-2}  \sin^2\omega_c
z)^{-1/2}$ and $w_y=\alpha_2(c^{-2}  \cos^2\omega_c z+b^{-2}  \sin^2\omega_c
z)^{-1/2}$, with those from the numerical simulation of
equation~(\ref{NSE1}), we find an excellent agreement as shown in
figure~\ref{width}. The method of numerical simulation used here
is the split-step Fourier method~\cite{Agrawal-book-2001-ssfm} by
using equation~(\ref{solution}) as the
input beam at $z=0$.
% where $b=4.26,c=2.13,P=127.32,\Theta=0.052,\beta_0=0$ and $\theta=0$.
\section{Rotation mode}
Now we pay attention to the rotation of the spiralling elliptic breathers in the media with the linear anisotropy. To this end, we should transform the expression of the elliptic optical beam, equation~(\ref{solution}), to its standard elliptic form through the rotation of coordinates by the angle $\vartheta$. We can obtain that $\tan(2\vartheta)=\frac{\gamma_{xy}}{\gamma_{xx}-\gamma_{yy}}$, where $\gamma_{xx}=(b^{-2}\cos^2\beta+c^{-2}\sin^2\beta)/\alpha_1^2,\gamma_{yy}=(b^{-2}\sin^2\beta+c^{-2}\cos^2\beta)/\alpha_2^2$ and $\gamma_{xy}=2\sin\beta\cos\beta(b^{-2}-c^{-2})/(\alpha_1\alpha_2)$. One of the semi-axes of the standard elliptic optical spot is
\begin{equation}\label{oneaxis}
  w_b=\sqrt{\frac{1}{\gamma_{xx}\cos^2\vartheta+\gamma_{yy}\sin^2\vartheta+\gamma_{xy}\sin\vartheta\cos\vartheta}},
\end{equation}
the other is
\begin{equation}\label{theother}
  w_c=\sqrt{\frac{1}{\gamma_{xx}\sin^2\vartheta+\gamma_{yy}\cos^2\vartheta-\gamma_{xy}\sin\vartheta\cos\vartheta}}.
\end{equation}
From equations~(\ref{oneaxis}) and (\ref{theother}), it is found that the two semi-axes of the elliptic optical beam vary with the propagation distance $z$ for the general case of $\alpha_1\neq \alpha_2$ and $b\neq c$. No spiraling elliptic solitons exist in the model (\ref{NSE1}), but only spiraling elliptic breathers exist, the semi-axes of which vary with $z$ periodically. For the case of $\alpha_1=\alpha_2$, from equations~(\ref{oneaxis}) and (\ref{theother}) we can obtain $w_b=\alpha_1 c,w_c=\alpha_1 b$, which is the case of Ref.\cite{Desyatnikov-prl-10} that the spiraling elliptic solitons can form in the saturable nonlinear media with linear isotropy, rotating with a constant angular velocity. For the case of $b=c$, we can obtain $\vartheta(z)=0$, that is, the two semi-axes of the elliptic solitons lie on the $x$ and $y$ axes all the time. There is no rotation. It is the case of Ref.\cite{guo-joa-1999}, where an elliptic
self-trapping beam for the extraordinary light is found to exist in the uniaxial crystal and the semi-axes of the elliptic optical beam lies on the principal plane of the uniaxial crystal.

\begin{figure}[htb]
\centerline{\includegraphics[width=4.5cm]{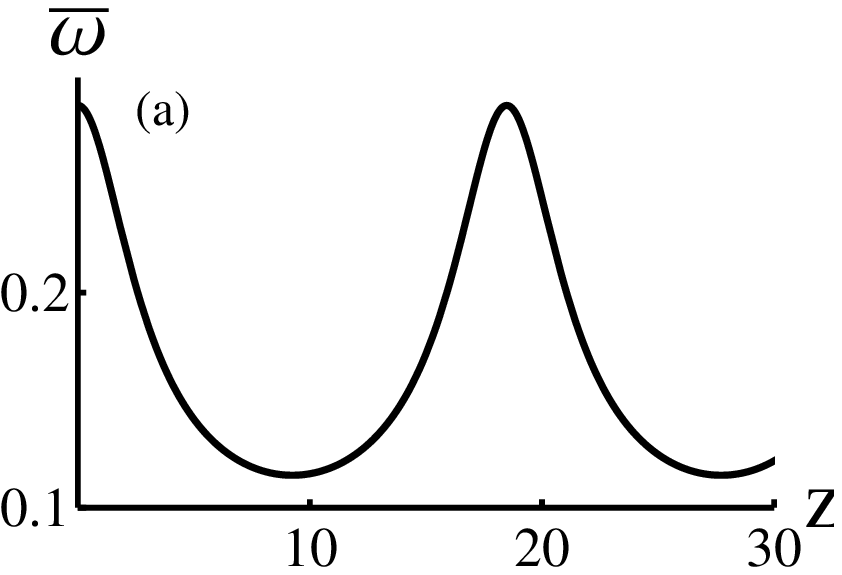}
\includegraphics[width=4.5cm]{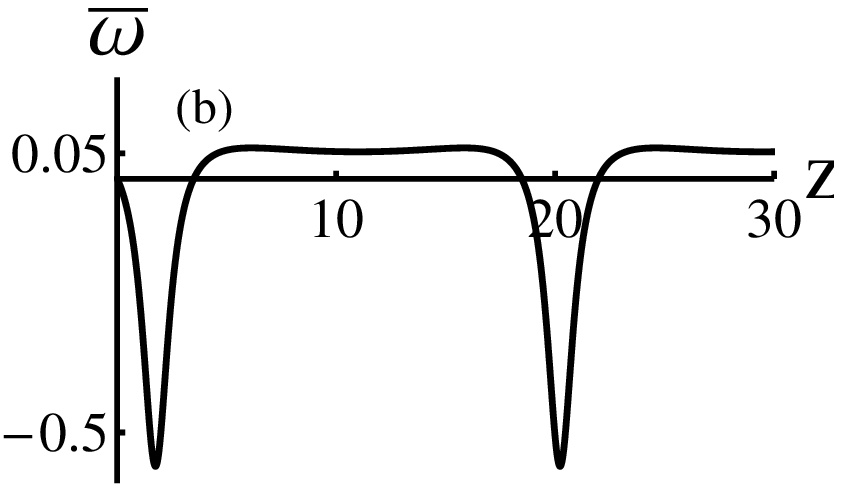}}
\caption{Variation of rotational angular velocity with propagation distance. The parameters are the same as those in figures~\ref{across} and \ref{width} for (a), and the parameters for (b) are $b=4.26,c=2.13,P=127.32,\Theta=0.052,\alpha_1=2.5,\alpha_2=1.0,\beta_0=73^\circ$.}\label{jiao1}
\end{figure}

The angular velocity of the optical beam in the $xyz$-coordinate frame can be obtained as
\begin{equation}\label{jiaosudu}
  \varpi=\frac{d\vartheta}{dz}=\frac{f_1(b,c,\omega,\alpha_1,\alpha_2,z)}{f_2(b,c,\omega,\alpha_1,\alpha_2,z)},
\end{equation}
where $f_1(b,c,\omega,\alpha_1,\alpha_2,z)=\alpha_1\alpha_2(b^2-c^2)\omega[(\alpha_1^2+\alpha_2^2)(b^2-c^2)+(\alpha_1^2-\alpha_2^2)(b^2+c^2)\cos2(\omega z+\beta_0)]$ and $f_2(b,c,\omega,\alpha_1,\alpha_2,z)=2(\alpha_1^2b^2-\alpha_2^2c^2)^2\cos^4(\omega z+\beta_0)+2(\alpha_2^2b^2-\alpha_1^2c^2)^2\sin^4(\omega z+\beta_0)+[\alpha_1^2\alpha_2^2(b^4+c^4)+(\alpha_1^4-4\alpha_1^2\alpha_2^2+\alpha_2^4)b^2c^2]\sin^22(\omega z+\beta_0)$. The angular velocity $\varpi$ is a periodic function of $z$ but not a constant, which is shown in figure~\ref{jiao1} (a).
We can prove that $f_2\geq0$ in equation~(\ref{jiaosudu}). If $f_1$ is always positive or negative at any propagation distance $z$, the spiraling elliptic breathers will rotate anticlockwise or clockwise. Mathematically, positive $f_1$ at any $z$ only requires that the minimum of $f_1$ is positive, i.e., $\min(f_1)\geq0$. Similarly, negative $f_1$ at any $z$ only requires that the maximum of $f_1$ is negative, i.e., $\max(f_1)\leq0$. From this we can obtain the criteria for the rotation of the spiraling elliptic breathers as $\min(\rho,1/\rho)\leq\rho_M\leq\max(\rho,1/\rho)$ but $\rho_M\neq1$, where $\rho=b/c$, and $\rho_M=\alpha_1/\alpha_2$ is the parameter of the linear anisotropy of the media.
The larger is the difference between $\rho_M$ and $1$, the larger is the linear anisotropy of the media.
 We find that $f_1$ is always positive when $\min(\rho,1/\rho)\leq\rho_M\leq\max(\rho,1/\rho)$, that is, the spiraling elliptic breather will rotate anticlockwise when the linear anisotropy of the media is small enough. As a matter of fact, we have used the positive $\omega$ here, which makes $f_1$ always positive. Of course, we can also use a negative $\omega$, then the spiraling elliptic breather will rotate clockwise when $\min(\rho,1/\rho)\leq\rho_M\leq\max(\rho,1/\rho)$. From equation~(\ref{angular velocity}), it can be found that the sign of $\omega$ depends on sign of $\sigma$. But the signs of $\sigma$ have no effects on equations~(\ref{1}) and (\ref{2}).
 Taking parameters $b=4.26,c=2.13,\alpha_1=1.0$ and $\alpha_2=1.3$ as an example, we can find the parameter $\rho_M$ satisfy $1/\rho<\rho_M<1$, then can predict the spiraling elliptic breathers will rotate anticlockwise, which is confirmed by figures~\ref{across} and \ref{jiao1} (a).

\begin{figure}[htb]
\centerline{\includegraphics[width=8.0cm]{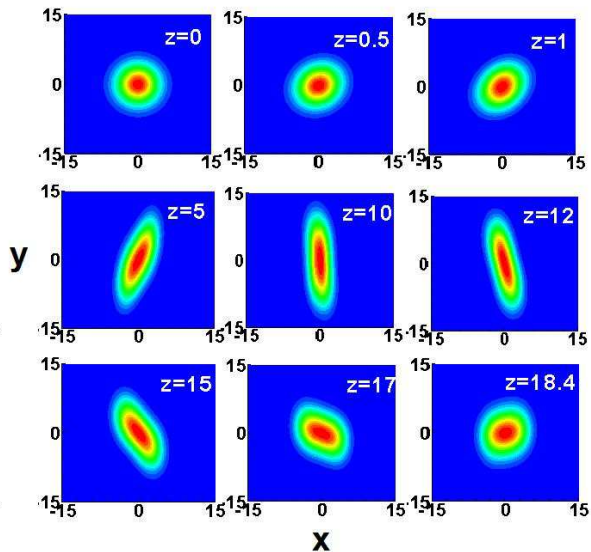}}
\caption{(color online) Evolution of spiralling optical breather in one period. The parameters are the same as those in figure~\ref{across} except for $\alpha_1=1,\alpha_2=2$ so that $\rho_M=1/\rho$.}\label{nooam}
\end{figure}

 The OAM of the optical beam in the $xyz$-coordinate frame is obtained as $m=\mbox{Im}\int\!\!\!\int\varphi^*(x\frac{\partial\varphi}{\partial y}-y\frac{\partial\varphi}{\partial x}){\rm d}x{\rm d}y=\frac{P\Theta f_1(b,c,\omega,\alpha_1,\alpha_2,z)}{4\omega\alpha_1\alpha_2(b^2-c^2)}$. If we set
$f_1(z=0)=0$, the spiraling breathers will have no initial OAM. Meanwhile, by setting $\min(f_1)=0$ or $\max(f_1)=0$, the spiraling breathers without the initial OAM will rotate anticlockwise or clockwise. Then we obtain that the spiraling breathers without the initial OAM rotate anticlockwise when $\rho_M=1/\rho, \beta_0=0$, as is shown in figure~\ref{nooam}, and rotate clockwise when $\rho_M=\rho$ and $\beta_0=\pi/2$. Such rotation of the spiraling elliptic breathers without the initial OAM in the media with the linear anisotropy is different from the rotation resulting from the initial OAM in the isotropy media~\cite{Rotschild-prl-05,liang-pra-2013}.

From the expression of the spiraling elliptic breathers, equation~(\ref{solution}), we can obtain the period of the evolution as $T=\pi/\omega$. Then the period of the spiraling elliptic breather in figure~\ref{nooam} is about $18.4$, where $\omega=0.17$, but we find that the shape of the the spiraling elliptic breather after a period is not the same as the input one at $z=0$. It is mainly because that the exact solutions of NLSE (\ref{NSE1}) are not the Gaussian function, the variational approach will bring in deviations if we take the Gaussian trial solution like equation~(\ref{trial}). The deviations will be expected to disappear if we use the numerical iterative solution of NLSE (\ref{NSE1}) as the input beam at $z=0$. Nonetheless, we can correctly predict the behaviors of the rotation and libration of the spiraling elliptic breathers by using the variational approach.

 \section{Pendulum-like libration mode}
\begin{figure}[htb]
\centerline{\includegraphics[width=7.0cm]{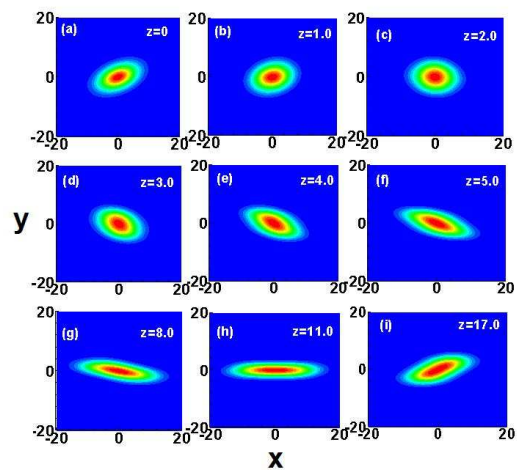}}
\caption{(color online) Evolution of the spiralling optical breather. The parameters are the same as those in figure~\ref{jiao1} (b).}\label{Evolution}
\end{figure}

 If we increase the linear anisotropy of the media so that $\rho_M>\max(\rho,1/\rho)$ or $\rho_M<\min(\rho,1/\rho)$, the angular velocity $\varpi$ will change its sign during the propagation of the spiraling elliptic breathers, that is, the direction of the rotation of the spiraling elliptic breathers will change, and the libration will appear, librating back and forth like the pendulum, which is shown in figures~\ref{jiao1} (b) and \ref{Evolution}. The spiraling breather rotates clockwise from $z=0$ to $z=3$, then changes the rotational direction and rotate anticlockwise until $z=17$. The critical angle, at which the rotation direction changes, can be found as
 \begin{equation}\label{crtitical angle}
 \tan(2\vartheta_{c})=\frac{1-\rho^2}{\sqrt{\rho^2-1-\rho^4+\rho^2\rho_M^2}}.
 \end{equation}
When $\rho=2,\rho_M=2.5$, the critical angle can be obtained to be $-25^\circ$, which agrees well with that of the numerical simulation, as is shown in  figure~\ref{Evolution} (d).

\begin{figure}[htb]
\centerline{\includegraphics[width=8.0cm]{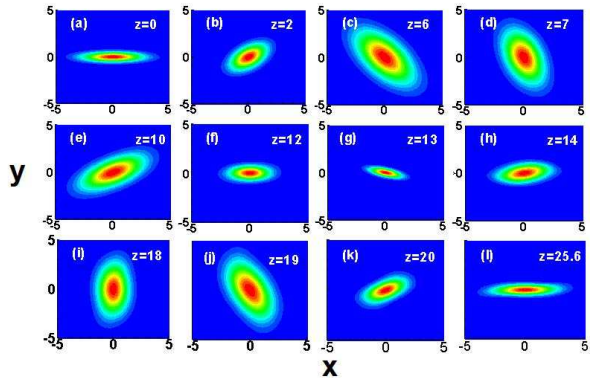}}
\caption{(color online) Evolution of the spiralling optical breather in one period. The parameters are the same as those in figure1 (a) of \cite{liang-pra-2013} except that the diffraction coefficients in the $x$ and $y$ directions of equation~(2) of \cite{liang-pra-2013} are $\frac{5\sqrt{2}}{8}$ and $\frac{\sqrt{2}}{4}$.}\label{nonlocal}
\end{figure}

It can be proved that the OAM is not conserved by the NLSE with unequal diffraction coefficients, so the libration of the spiraling elliptic breathers is universal in the nonlinear media with the linear anisotropy if the anisotropy is large enough.
For example, apart form the saturable nonlinear media, we can simply explore the libration of the spiraling elliptic breathers in nonlocal nonlinear media with the linear anisotropy. We simulate the nonlocal nonlinear Schr\"{o}dinger equation $ i\frac{\partial\varphi}{\partial z}+\alpha_1^2\frac{\partial^2\varphi}{\partial x^2}+\alpha_2^2\frac{\partial^2\varphi}{\partial y^2}+\varphi\int\int R(x-x',y-y')|\varphi(x',y')|^2dxdy=0$
taking the optical beam (\ref{solution}) as an input beam with the response function of the media $R=1/(2\pi w_m^2)\exp[-(x^2+y^2)/(2w_m^2)]$, where the parameters, $w_m=15,b=2,c=1,P=1.25\times10^5, \Theta=0.374$,
are the same as those in the figure1 (a) of Ref.\cite{liang-pra-2013}. But instead of $\alpha_1=\alpha_2=\frac{\sqrt{2}}{2}$ in \cite{liang-pra-2013}, here we use the diffraction coefficients $\alpha_1=\frac{5\sqrt{2}}{8}$ and $\alpha_2=\frac{\sqrt{2}}{4}$ to make sure that the parameter of the linear anisotropy, $\rho_M=2.5$, and $\rho=b/c=2$ are the same as those in figure~\ref{Evolution} of this paper. The evolution of the spiraling elliptic breather is shown in figure~\ref{nonlocal}, where the libration happens too. The spiraling breather rotates anticlockwise from $z=0$ to $z=6$, then changes the rotational direction and rotate clockwise until $z=13$, then changes the rotational direction again, and rotate anticlockwise until $z=19$, then changes the rotational direction again, and rotate clockwise until $z=25.6$.  The shape of the spiraling elliptic breather at $z=25.6$ after one period is almost the same as that of the input beam at $z=0$. The larger the degree of nonlocality is, the less the difference of the shapes between the beam after one period and the input beam is. The details about the rotation and libration of the spiraling elliptic breathers in nonlocal nonlinear media with the linear anisotropy will not be reported herein.

\section{Conclusion}

We have introduced a class of spiraling elliptic breathers in saturable nonlinear media with linear anisotropy. The breathers have the elliptic light spots, and have two different kinds of evolution modes, that is, the rotation mode and the pendulum-like mode. The spiraling elliptic breathers can rotate even though
they have no initial orbital angular momentum. Due to the anisotropy of the media, the angular velocity is not a constant but a periodic function of the propagation distance. When the anisotropy of the media is large enough, the spiraling elliptic breathers will librate like the pendulum. The spiraling elliptic breathers are universal in the nonlinear media with the linear anisotropy, such as the saturable nonlinear media, the nonlocal nonlinear media and so on. The predictions of the rotation and libration of the spiraling elliptic breathers by the variational approach are confirmed by the numerical simulation.

\section*{ACKNOWLEDGMENTS}\label{ACKNOWLEDGMENTS}

This research was supported by the National Natural Science
 Foundation of China (Grant Nos. 11074080, 11274125), and the Natural Science Foundation of Guangdong Province of China (Grant No. S2012010009178).

%\begin{thebibliography}{99}
%%% Do not include separate BibTeX files; if BibTeX is used,
%%% paste the output (contents of .bbl file) here.
%
%\bibitem{revtex-au} \url{https://authors.aps.org/revtex4/}.
%\bibitem{osastyle} \url{http://www.opticsinfobase.org/submit/style/jrnls_style.cfm}.
%
%\end{thebibliography}

\end{document}